\documentclass{aims}

\usepackage{txfonts,cite}
\def\typeofarticle{Research article}
\def\currentvolume{17}
\def\currentissue{5}
\def\currentyear{2020}

\def\ppages{xxx--xxx}
\def\DOI{}
\def\Received{04 July 2020}
\def\Accepted{13 August 2020}
\def\Published{August 2020}

\numberwithin{equation}{section}

\usepackage[pagewise]{lineno}

\emergencystretch=\maxdimen
\begin{document}

\title{The dynamics of COVID-19 spread: evidence from Lebanon}

\author{%
  Omar El Deeb\affil{1,2,}\corrauth 
and
  Maya Jalloul\affil{3}
}

% \shortauthors is used in copyright information in the end of the paper
\shortauthors{the Author(s)}

\address{%
  \addr{\affilnum{1}}{Lebanese University, Faculty of Science, Hadath, Lebanon}
  \addr{\affilnum{2}}{Lebanese International University, Mathematics and Physics Department, Beirut, Lebanon}
  \addr{\affilnum{3}}{Lebanese American University, Department of Economics, Beirut, Lebanon}}

% corresponding author
\corraddr{Email: omar.deeb@liu.edu.lb; Tel: +9615411678.
}

\begin{abstract}
We explore the spread of the Coronavirus disease 2019 (COVID-19)
in Lebanon by adopting two different approaches: the STEIR model, which is a modified SEIR model accounting for the effect of travel, and a repeated iterations model. We fit available daily data since the first diagnosed case until the end of June 2020 and we forecast possible scenarios of contagion associated with different levels of social distancing measures and travel inflows. We determine the initial reproductive transmission rate in Lebanon and all subsequent dynamics. In the repeated iterations (RI) model we iterate the available data of currently infected people to forecast future infections under several possible scenarios of contagion.
In both models, our results suggest that tougher mitigation measures would slow down the spread of the disease. On the other hand, the current relaxation of measures and partial resumption of international flights, as the STEIR reveals, would trigger a second outbreak of infections, with severity depending on the extent of relaxation. We recommend strong institutional and public commitment to mitigation measures to avoid uncontrolled spread.
\end{abstract}


\maketitle

\section{Introduction}

The Coronavirus disease 2019 (COVID-19) has been widely spreading
worldwide since it appeared in the city of Wuhan, China towards the
end of December, 2019. The World Health Organization classified the
spread as a pandemic in March 2020 \cite{WHO}. Europe and the United
States of America have endured the severest repercussions in terms of number of infections and deaths. The
US cases amounted to more than one fourth of total global infections by June 2020 \cite{worldometer2020}. Governmental and institutional reactions and measures varied across countries with respect to the time of introduction of social distancing measures (SDM, henceforth) and with respect to their degree of severity. In spite of some governments being slower in adopting mitigation measures and endorsing the epidemiological concept of herd immunity \cite{NYT2020} to create a resistance to the contagion in the long run at the expense of short term losses while keeping the economy functional, the majority adopted SDM's that reached countrywide lockdowns. 

A considerable amount of research has been carried out focusing on the dynamics and extent of the pandemic in different countries notably in the countries that witnessed the first cases \cite{Li2020,Zhou2020,Biswas,Yang, Fanelli2020}.
In comparison with the most recent deadly mass pandemic of the "Spanish flu" that hit the world after World War I during the years of 1918--1919 and recurred in two waves, and caused the death of tens of millions of people \cite{Spanish flu, Spanish flu III}, the extent of the spread of COVID-19 has been far less. The question of the containment of COVID-19 and preventing its spread and expansion into a similar deadly pandemic is a key motivation for the study of various models
that describe, simulate and forecast epidemics and dynamics of infections under different reproductive
rates and mitigation measures.

In Lebanon, the spread of COVID-19 coincided with a period of political turmoil, few months of popular uprising and economic collapse finally depicted by the default on debts in early March 2020~\cite{Guardian2020}. 

\paragraph{Lebanon SDM}
SDM  were adopted at a relatively early stage in Lebanon. 
The first confirmed case was recorded on February 21, 2020, and consequently all educational institutions were closed starting the first of March. This was followed by a resolution of "Public Mobilization" and ban of public gatherings that imposed closure of all places of worship, shops, restaurants, etc. except for grocery stores and drugstores on March 22 \cite{MOPH2020}. Another subsequent measure consisted in constraining vehicles' mobility to alternating between odd- and even-ending plate numbers, while fully prohibiting mobility on Sundays. Gradual and partial relaxation started on May 10, but the deep economic crisis helped in slowing down social activity despite the easing of measures. The government decided to open the airport with one-fifth of its normal capacity starting July 1st, after a period of limited returns for Lebanese people living abroad between April 8 and June 30.

The source of the first infection was documented before the international travel ban to be from a traveler coming from Iran \cite{Iran1} where the spread of the virus had started early on \cite{Iran2}. However, it has been discussed that many of the following first cases were transmitted by travelers coming from the Vatican city during the early stages of the pandemic spread there \cite{Gavlak}. The efficacy of SDM's can be examined by discerning the daily rate of infection as shown in the daily data of the first 130 days \cite{DRM2020}.  Our results were obtained in relation to data available until June 30, 2020.

\paragraph{Literature and methodology}
Models exploring contagion and particularly spread of infections were developed and extensively studied in various fields of mathematics, physics, economics and epidemiology \cite{Boccaletti2006,Pastor2015,Batista,LOU2020, Sun, Complexity}. Mitigation measures have been shown to be effective in reducing and slowing down the spread of infection by \cite{Anderson2020,Mcloskey2020, Kraemer2020}. 

The SEIR model \cite{Atkeson2020,SEIR2,SEIR3} has been widely employed to investigate contagion of diseases including that of COVID-19 recently. In this paper, we establish a novel STEIR model, which consists of modifying the SEIR model to take into account the effect of travel.  We use the STEIR and a repeated iterations model presented in \cite{Perc2020} (henceforth the RI model) and implement it on the daily data of the Coronavirus in Lebanon provided mainly by the Ministry of Public Health \cite{DRM2020} over a duration of 130 days.

\begin{figure}[h]
\hspace*{-1cm}
\begin{center}
\includegraphics[scale=0.5]{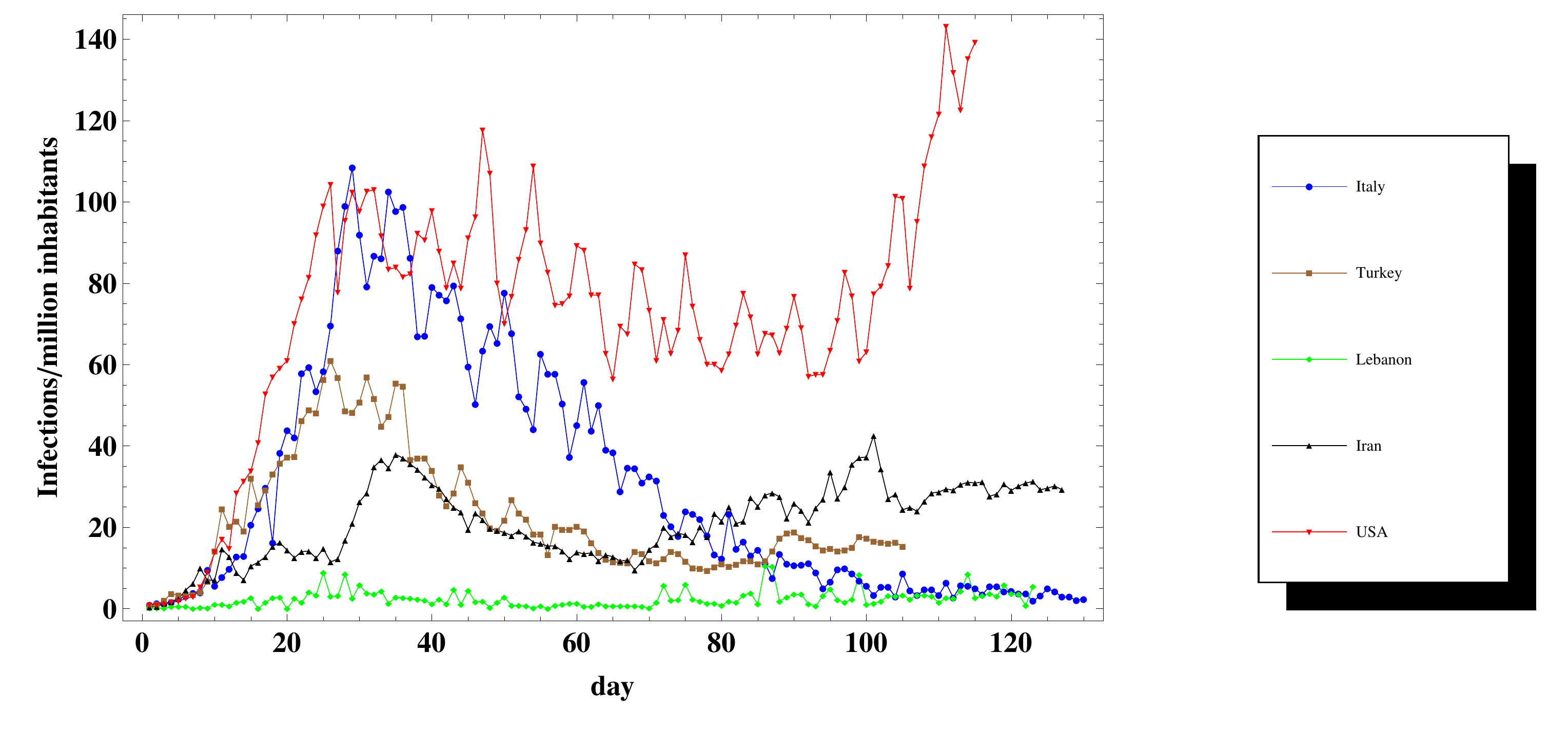}
\caption{Daily infections (per million) in Lebanon, Turkey, Iran, Italy and the USA. In comparison, Lebanon relatively maintained a low infection rate during the first 120 days.}  \label{fig:1}
\end{center}
\end{figure}

The structure of the paper is as follows. In section 2, we present the two models. Results and forecasts are discussed in section 3 and section 4 concludes the paper.

\section{Theoretical framework}

We describe here two models that we use to forecast the path of daily cases and to measure the extent of the epidemic in Lebanon.
Following an abrupt rise in infected cases at the start, the rate has fallen to a significantly low level after implementation of severe SDM, in comparison with other regional and country-level data \cite{ourworldindata2020} before slowly rising again after partial relaxation. Figure \ref{fig:1} illustrates the progression of the number of daily infections over time in different countries: Lebanon, Turkey, Iran, Italy and the USA. Day $0$ represents the date of the first reported infection. Lebanon has a relatively low number of infections per capita, despite the fact that the first case was recorded relatively early. The proposed models accommodate for the actual data and allow for future predictions.

\subsection{STEIR model}

The  spread of epidemics can be studied by the prominent SIR model first introduced by Kermack and McKendrick \cite{Kermack1927}, that consists of dividing the population into different compartments and investigating the contagion of the disease by examining the rate of change of the sizes of these groups. Later versions were developed to accommodate for different settings and assumptions \cite{May1991, Capasso1993, Vynnycky2010}. The STEIR model that we use is described as follows.   
Consider a population $N$ initially normalized to size~$1$ and divided into
five categories of individuals: susceptible $S$, incoming travellers $T$, exposed $E$, infectious $I$ and
removed (through recovery or death) $R$ \footnote{The exposed individuals represented by $E$ are those who are infected but not yet infectious, as defined in \cite{Wang2020}. Recent studies reveal that during incubation period, individuals are not infectious, and infectivity starts on an average of only 1-2 days before symptoms \cite{Anderson2020}.}.
 The cumulative number of cases is provided
by $C=I+R$, since each recorded case would at a given time be either infectious
or recovered or dead. The rates of change of these categories
are given by $\frac{dS}{dt},\frac{dT}{dt}, \frac{dE}{dt},\frac{dI}{dt}$ and $\frac{dR}{dt}$
respectively. The model is formally characterized using the following differential equations:

\begin{equation}
\frac{dS}{dt}=-\beta_{t}\frac{S}{N}I +(1-\theta) \tau 
\end{equation}

\begin{equation}
\frac{dE}{dt}=\beta_{t}\frac{S}{N}I-\sigma E
\end{equation}

\begin{equation}
\frac{dI}{dt}=\sigma E-\gamma I +\theta \tau
\end{equation}

\begin{equation}
\frac{dR}{dt}=\gamma I
\end{equation}

\begin{equation}
\frac{dN}{dt}=\tau N
\end{equation}

\noindent with $\beta_{t}=R_{t}\gamma$ where $\beta_{t}$ is the rate at which
infected individuals come across others (the $\frac{S}{N}$ susceptibles),
$\sigma$ is the rate at which exposed individuals become infected
and it is associated to the mean incubation period; $\gamma$ is the rate
of exit by recovery or death per-day and it is associated to the average
illness period. In this sense, $R_{t}=\frac{\beta_{t}}{\gamma}$ is the ratio of meeting rate to exit rate;
it determines the transmission from susceptible to infected and is
a proxy for social distancing measures. $\tau$ is the daily rate of incoming travellers relative to the total population $N$ and it contributes to both, the infected $I$ and the susceptible $S$.  $\theta$ is the average infection rate among the travellers.
 The population size $N$ slightly increases due to the influx of travellers whose relative contribution is more apparent in the increase in the number of infections $I$. We neglect the natural changes in the population as the time scale of the epidemic is much faster than that of demographic changes \cite{Wang2020}.

There are several ways to model $R_{t}$. It can be taken as a constant parameter in some contexts, or as a time dependent function as  in \cite{Atkeson2020}. Here, we introduce a novel parameterization for $R_{t}$. Using data available from the first 130 days in Lebanon, we parameterize  $R_{t}$ by the step function
\begin{equation}
R_{t}=
\begin{cases}
R_{0} & 0<t<t_{1}\\
R_{1} & t_{1}<t<t_{2}\\
R_{2} & t_{2}<t<t_{3}
\end{cases}
\end{equation}

after the application of SDM, before any relaxation, then after partial relaxation at time $t_{2}$ respectively. $t_{3}=130$ days, corresponding to the last day of used data.  When the measures are changed after this period of SDM, $R_{t}$ is parameterized as follows:

\begin{equation}
R_{t}=\begin{cases}
R_{0} & 0<t<t_{1}\\
R_{1} & t_{1}<t<t_{2}\\
R_{2} & t_{2}<t<t_{3}\\
R_{3}& t>t_{3}
\end{cases}
\end{equation}

\noindent where $R_{0}$, $R_{1}$, $R_{2}$ and $R_{3}$ are constant parameters that
depend on the severity of measures and commitment to those measures. $R_{0}$ is the reproductive transmission rate of the disease in the initial phase, while $R_{1}$ and $R_{2}$ are the reproductive transmission factors under strict and relaxed SDM respectively. $R_{3}$ represents the potential future rate under different possible relaxation schemes. 
Using the estimates in \cite{Wang2020}, we take $\sigma=\frac{1}{5.2}$
in relation to an average period of incubation of $5.2$ days. We use
$\gamma=\frac{1}{20}$ underlying an average period of recovery
(or death) of $20$ days. This value is in line with the reports of the World Health Organization that estimate the average time of recovery to be in the range of 2--3 weeks \cite{WHO2}.
We can determine $\tau$ from Eq (2.5), and we find that
\begin{equation}
\tau=\frac{\ln(\frac{N_{f}}{N_{i}})}{\Delta t}
\end{equation}
where $N_{i}$ is the population on April 8 and $N_{f}$ is the population on June 30, after the arrival of air passengers. The time $\Delta t$ is the number of days of incoming flights. This corresponds to $\tau=2.91\times10^{-5} /$day. From official data provided in \cite{DRM2020}, we also find that $\theta=0.0377$ by taking the ratio of all positive cases among arriving travellers to the total number of incoming travellers until June 30, 2020.

Assuming that initially no SDM are applied, $R_{0}$ represents the
transmission of disease with no mitigation measures. In \cite{Wang2020}
they adopt $R_{0}=3.1$, while in \cite{Remuzzi2020} they take values
between 2.76 and 3.25, and \cite{Atkeson2020} considers different
values between 3 and 1.6. Recent studies reveal that $R_{0}$ of COVID-19 can assume higher values up to 5.7 and 6.47 according to data analyzed  from China \cite{CDC2020, Tang2020}.
After introduction of measures, $R_{t}$ can assume values less than 1
in case extremely severe mitigation measures are applied \cite{Dushov2014}.
It was well established that COVID-19 has higher $R_{0}$ than other
infections like SARS \cite{Liu2020}. The estimation of $R_{0}$ and
${R_{t}}$ is essential for forecasting the spread, but their determination
depends on the available data and the accuracy of the reporting of
initial cases and dates.

We assume that the initial value of $I$ is $I_{0}=\frac{1}{6M}$,
in line with the first initial case in Lebanon reported on February
21 and a gross population of 6 million inhabitants. We take $E_{0}=12I_{0}$
to account for the fact that the initial case registered had been
in contact with many people on a flight from Iran which raises the number of initially exposed people. The passengers were unprotected as they were not aware of the presence of an infection among them until late during the flight \cite{IranFlight}. To account for this fact, we took $E_{0}$ to be triple the value used by \cite{Atkeson2020} where they consider $I_{0}=\frac{1}{10M}$
with $33$ initial cases in the United States whose population stands at
around 330 millions, and $E_{0}=4I_{0}$ given $132$ individuals were
initially carrying but not contagious, acknowledging the considerable
uncertainty related to initial cases in the US. This entails some uncertainty in the initial conditions of the spread due to the lack of readily available data.

By fitting the values of the reproductive number parameters in our model, we find that $R_{0}=5.6$, $R_{1}=0.52$  and $R_{2}=1.1$ for $t_{1}=32$ days, $t_{2}=63$ days and $t_{3}=130$ days respectively, provide the best numerical prediction to account for the actual reported cases. We assume that relaxation of measures would increase the reproductive rate, but that public awareness and remaining measures will keep it well below its initial value. Accordingly, we simulate four possible future scenarios with $R_{3}=1.1$, $1.4$, $1.7$ and $2$ after $t_{3}=130$ days. Regarding $\tau$, we inspect four possible cases for $t>130$ days, upon the opening of Beirut airport, and the expected increase in the number of travellers entering the country. We simulate the effect of $\tau$ multiplied by $2$, $3$ and $4$ in comparison to the previous rate of arrivals, and its consequences on the cumulative number of infected cases.

\subsection{The RI model}

In this section, we adopt an alternative approach, the RI model, which exploits the available daily data of infections and recoveries to forecast progression of the disease.  Here we implement a variation of the repeated iterations method proposed in \cite{Perc2020}. We denote the active infected
daily values by $I_{i}$ where $i$ is the index of days and $i\in[1,n]$.
We use the last $m$ values of $I_{i}$ to determine the average
arithmetic gross rate in the last $m$ days as

\begin{equation}
G_{a}=\frac{1}{m}\sum_{i=n-m+1}^{n}\left(\frac{I_{i}}{I_{i-1}}-1\right)
\label{eq:13}
\end{equation}

\noindent  and the average geometric gross rate of the same set of data as

\begin{equation}
G_{g}=\left(\frac{I_{i}}{I_{i-m}}\right)^{\frac{1}{m}}
\label{eq:14}
\end{equation}

\noindent  $G_{a}$ and $G_{g}$ indicate the average rate of increase in the number of current infections during the last $m$ days, using arithmetic and geometric means. We forecast the progression of the number of infections accordingly as follows

\begin{equation}
I_{i+1}=I_{i}\left(1+G_{a}\right)
\end{equation}

\noindent  using the arithmetic gross infection rate $G_{a}$ or alternatively

\begin{equation}
I_{i+1}=I_{i}G_{g}
\end{equation}

\noindent  using the geometric gross infection rate $G_{g}$.

To account for the removal of active cases, we define the death and recovery rates by $p$ and $1-p$ respectively,  the number of days need for recovery by $h$  and the average number of days between infection and death by $d$. This implies that
on day $(i+1)$, the number of deaths will be $p\left(I_{i-d}-I_{i-d-1}\right)$
where $\left(I_{i-d}-I_{i-d-1}\right)$ represents the number of people who caught
the virus $d$ days ago. Similarly, the number of people recovered
would be proportional to the number of people who caught the virus
$h$ days ago and is given by $\left(1-p\right)\left(I_{i-h}-I_{i-h-1}\right)$.
 The values of $p$ and $h$ vary in the literature and in the available
data from the country under consideration. To simulate the available data from Lebanon, we take $p=0.04$, $h=20$ days and $d=24$ days.  $h$ is in accordance with the average period of recovery taken in the STEIR model used before, and $p$ and $d$ are discussed in available literature \cite{Perc2020}.

Given the aforementioned parameters, we define the recursive relation used in our forecast as follows

\begin{equation}
\left(I_{i+1}\right)_{net}=I_{i+1}-p\left(I_{i-d}-I_{i-d-1}\right)-\left(1-p\right)\left(I_{i-h}-I_{i-h-1}\right)
\end{equation}

\noindent  where $\left(I_{i+1}\right)_{net}$ is the net number of infected people on day $(i+1)$. In every iteration, we use the available data up to the day under consideration, estimate the number of infections on the next day, and reevaluate $G_{a}$ and $G_{g}$ accordingly, hence recursively predicting the progression of the rate and the number of active infected cases.  
We will consider $m=14$ previous
days, and forecast the next upcoming $14$ days as well. The interesting
feature of the RI model is that as new daily data is revealed, we update our daily future forecast, hence have a new $14$ day
future expectation every day.

\begin{figure}[t]
\hspace*{-2cm}
\begin{center}
\includegraphics[scale=0.5]{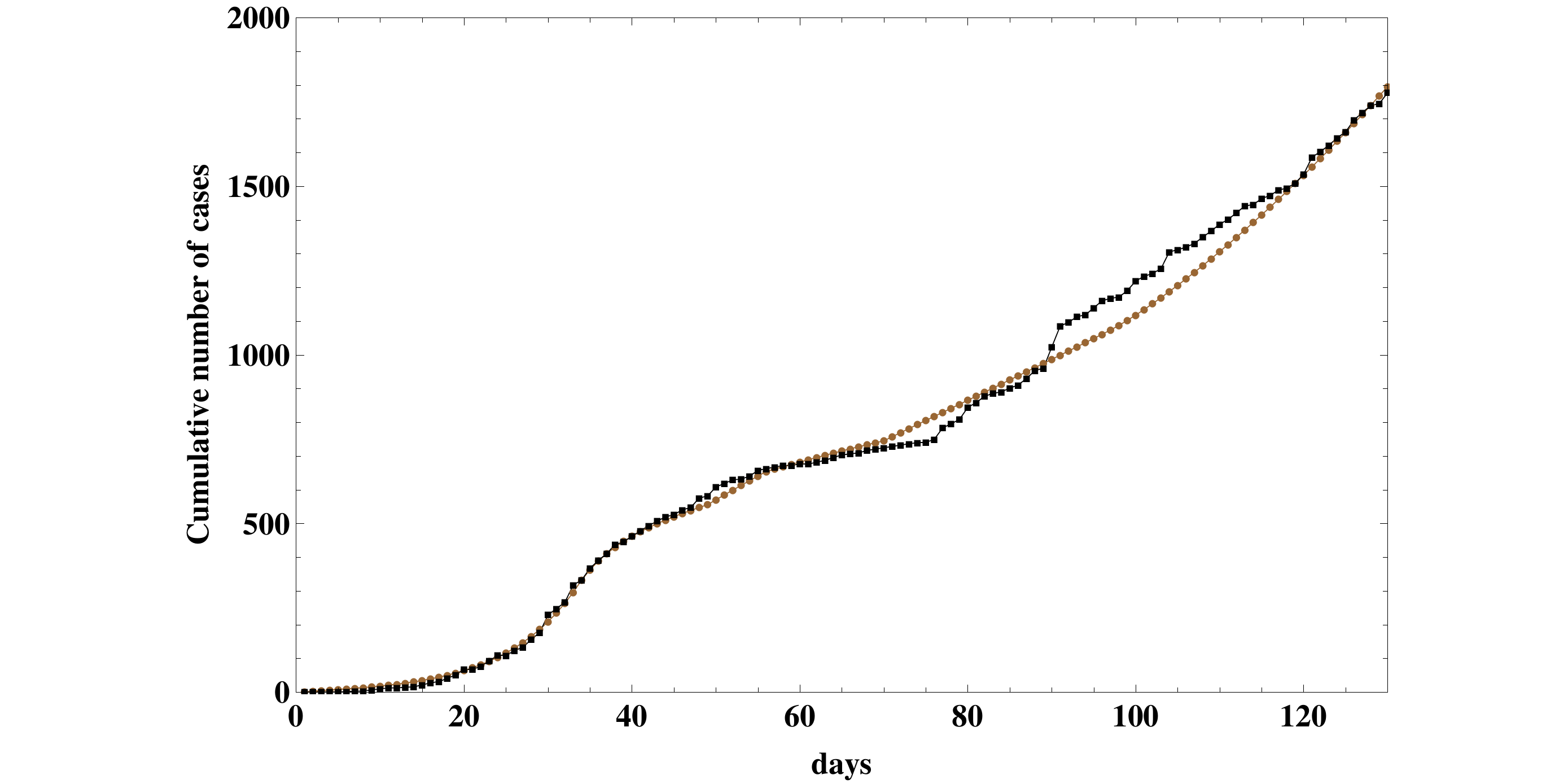}
\caption{Cumulative number of infections according to the STEIR model in Lebanon, with appropriate parameterization with $R_{0}=5.6$ for the first 32 days, $R_{1}=0.52$ for the next 63 days and $R_{2}=1.1$ until day 130. The actual cases are represented in black and the predicted cumulative infections are in brown.}\label{fig:2}
\end{center}
\end{figure}

The average gross rate $G$ (the arithmetic rate $G_{a}$ or the geometric rate $G_{g}$)  is a dynamic quantity and it is contingent on the promptness and severity of public policies of isolation, curfews and social distancing as well as on the degree of individual commitment and personal preventive efforts e.g. sanitation, use of face masks, etc. For this end, we consider several scenarios to predict the progression of $G$ in relation to SDM. Furthermore, we conduct the simulations using both $G_{a}$, $G_{g}$, the maximum value of $G$ attained in the previous $m$ days, and possible increases or decreases of those rates.

\section{Results}

\subsection{STEIR}

\begin{figure}[h]
\hspace*{-2cm} 
\begin{center}
\includegraphics[scale=0.5]{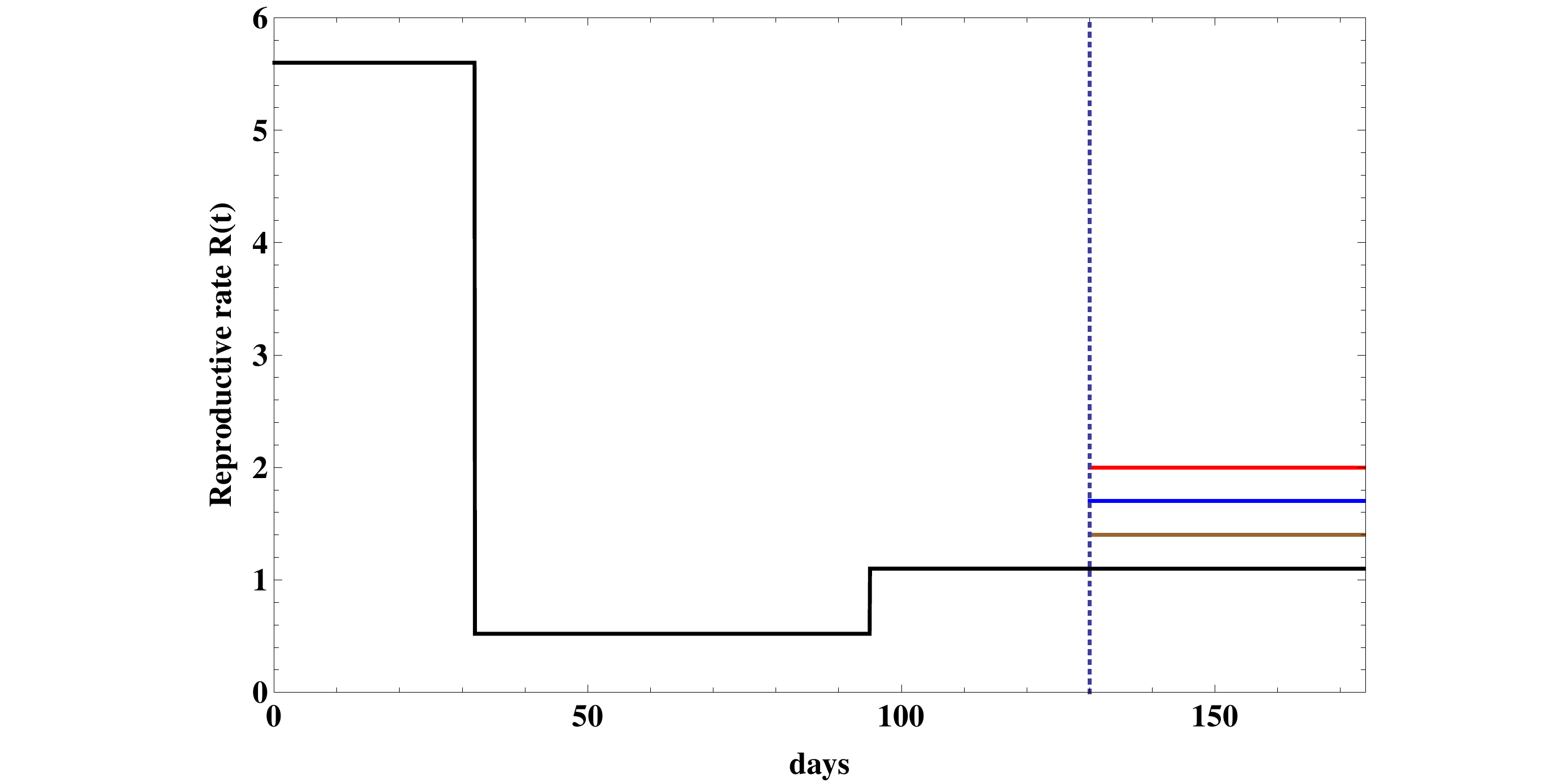}
\caption{The reproduction transmission factor $R(t)$ as a function of time $t$ in days. It starts at $R_{0}=5.6$ then falls down at $t=32$ days to $R_{1}=0.52$ after strong mitigation measures, then rises to $R_{2}=1.1$ at $t=90$ days. At $t=130$ days, we inspect four possible scenarios of $R_{3}$ corresponding to different levels of relaxation of mitigation measures, with numerical values equal to $1.1$, $1.4$, $1.7$ and $2$ in black, brown, blue and red respectively.  } \label{fig:3}
\end{center}
\end{figure}

In the parameterization of the STEIR model based on the cases reported in Lebanon, we find that the initial value of the reproductive transmission factor is remarkably high at $R_{0}=5.6$. However, it significantly decreases to $R_{1}=0.52$ at $t_{1}=32$ days, before rising moderately to $R_{2}=1.1$ at $t_{2}=90$ days following a partial relaxation of SDM. This parameterization provides an accurate fit of the registered cases for the first $130$ days. In addition to the absence of any SDM, a relatively high value of $R_{0}$ could be attributed to late reporting or under reporting of the early cases, hence the fast surge of registered results during the first few days of testing. $R_{0}$ here reflects the fast pace of spread of the disease in the early days before implementing social distancing measures. The rapid fall of the rate from $R_{0}$ to $R_{1}$ occurs after the implementation and the social commitment to mitigation measures, hence the rapid decrease in the number of daily infections and the slow increase in the cumulative number of infections. The significantly low rate of $R_{1}$ sharply diminishes the number of new infections, and the curve of the cumulative number of infections (Figure  \ref{fig:2}) starts flattening out slowly, before its new rise when the SDM are eased.

However, if the measures are relaxed at a time $t_{3}=130$ days, we expect an increase in the reproductive rate from $R_{2}$ to a another constant value $R_{3}$. The exact value of  $R_{3}$ will still depend on the extent of relaxation and the public commitment to SDM. We consider here four possible values of $R_{3}$ (Figure  \ref{fig:3}): continued partial mitigation measures with $R_{3}=R_{2}=1.1$, a weakly higher relaxation with $R_{3}=1.4$, while more public social interaction and moderate relaxation has $R_{3}=1.7$. Finally, we investigate $R_{3}= 2$ which corresponds to a wider relaxation of measures that yet coincides with self-awareness and individual efforts, hence preventing the return to the initial rate  $R_{0}$.
 
\begin{figure}[t]
\hspace*{-2cm}
\begin{center}
\includegraphics[scale=0.55]{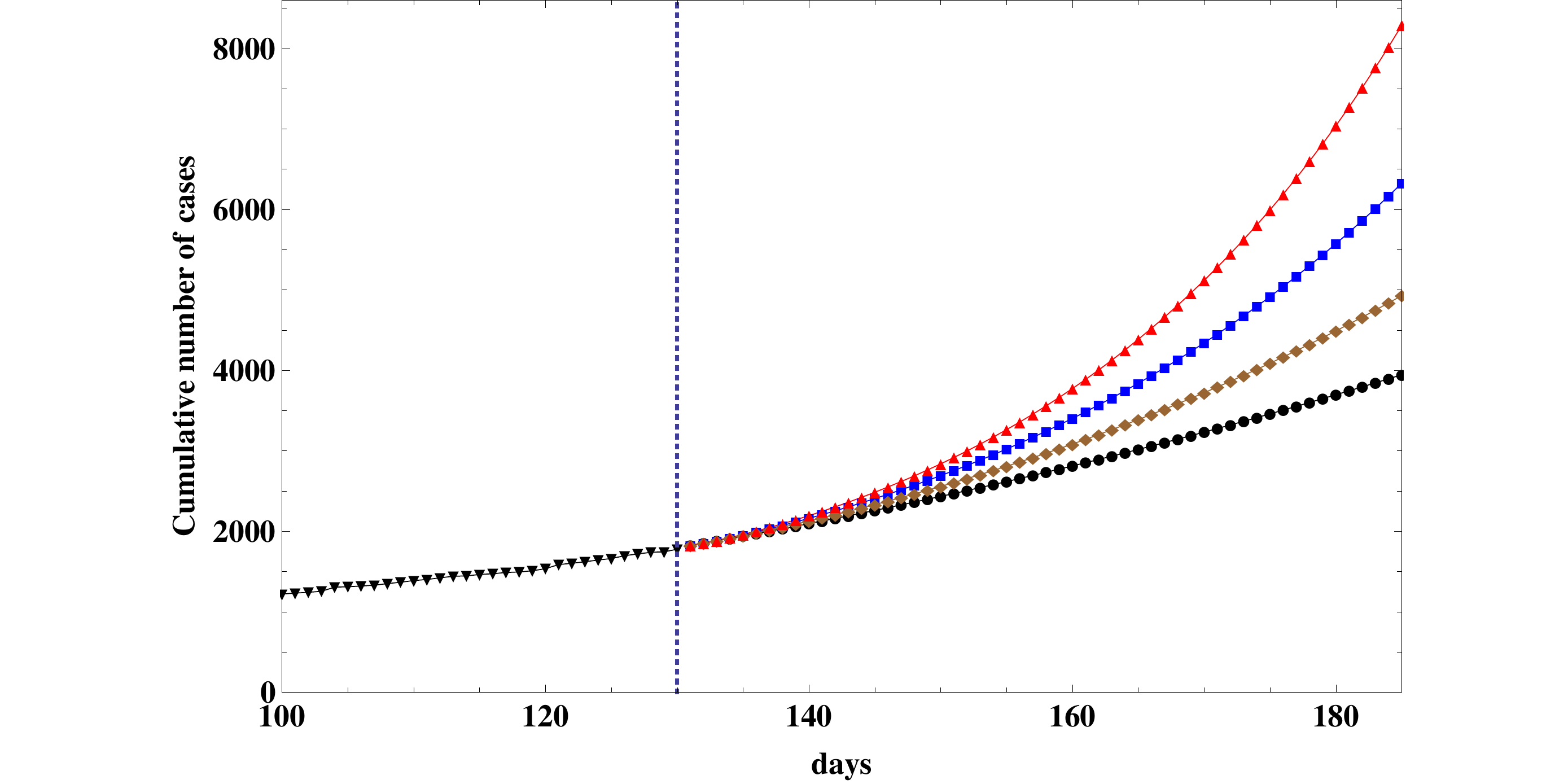}
\caption{Cumulative number of infections with different possible SDM relaxation scenarios after $t=130$ days. The figure represents a forecast with $R_{3}=1.1$, $1.4$, $1.7$ and $2$,  plotted in black, brown, blue and red respectively in correspondence with Figure \ref{fig:3}. It shows that the number of infections can still grow exponentially once the measures are relaxed. } \label{fig:4}
\end{center}
\end{figure}

\begin{figure}[t]
\hspace*{-2cm}
\begin{center}
\includegraphics[scale=0.5]{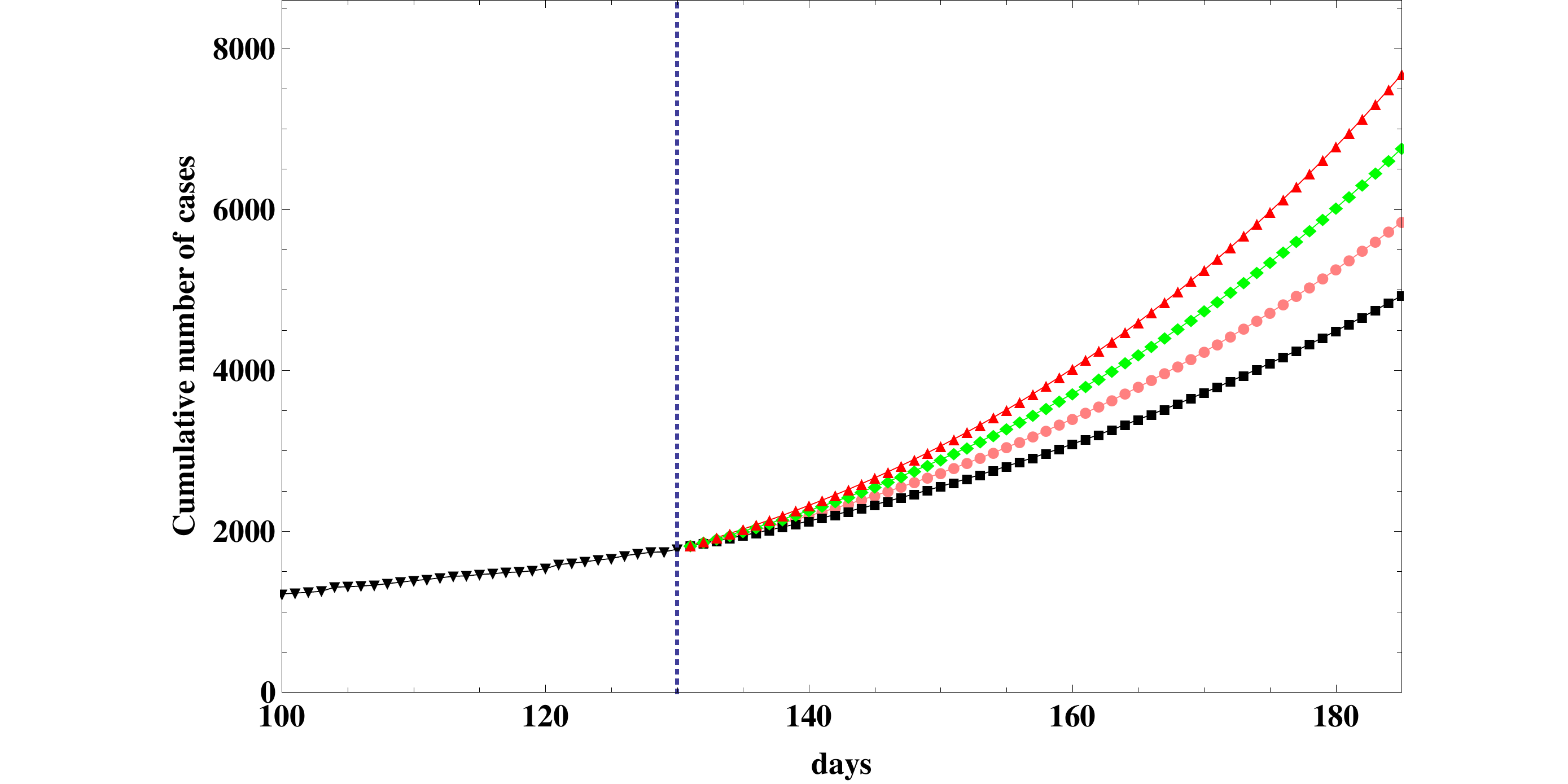}
\caption{Cumulative number of infections with four possible travel influx scenarios after $t=130$ days. The figure represents a forecast with $R_{3}=1.1$ and $\tau$ multiplied by $2$, $3$ and $4$,  plotted in black, pink, green and red respectively. The increase in international travel influx to the country leads to a rapid increase in the spread of the infection.} \label{fig:5}
\end{center}
\end{figure}

We find that the cumulative number of infections will rise again at a higher pace as depicted in Figure \ref{fig:4} even for the lowest increase in $R_{3}$. The number of cumulative infections could reach a total varying between $4491$ to more than $7042$ infections in a period of 50 days, depending on the extent of relaxation. A constant level of SDM of  $R_{2}$ would lead to a total of $3697$ infections over the same period. 
Moreover, we find that even for the low constant value of $R_{3}$, an increase in the flow of incoming travelers would drive the number of infectious cases upward, depending on the rate of travel over the period of the posterior $50$ days (Figure  \ref{fig:5}). More precisely, the simulation of the first $130$ days with absolute closure of the airport and no incoming travelers shows that the cumulative number of infections would have reached $948$  instead of the $1778$ registered cases on June 30.

This also suggests that the disease can rapidly spread again once the measures are relaxed, at a pace contingent on the degree of relaxation of the measures.  A combination of loosened measures and a higher rate of travel arrivals would trigger a stronger second phase of infection. These results are numerically specific to Lebanon, but they can be naturally generalized, and they suggest that a second outbreak of COVID-19 infections would be inevitable in absence or weakening of within as well as cross-country measures.

\subsection{RI model}

The arithmetic and geometric means proposed in equations (\ref{eq:13}) and (\ref{eq:14}) assume that all the SDM will remain unchanged over the $m$ subsequent days under consideration. Nevertheless, this might not be the case.  In order to take into account possible changes we consider the following scenarios that we simulate in Figure \ref{fig:6}:

\begin{figure}[t]
\hspace*{-2cm}
\begin{center}
\includegraphics[scale=0.6]{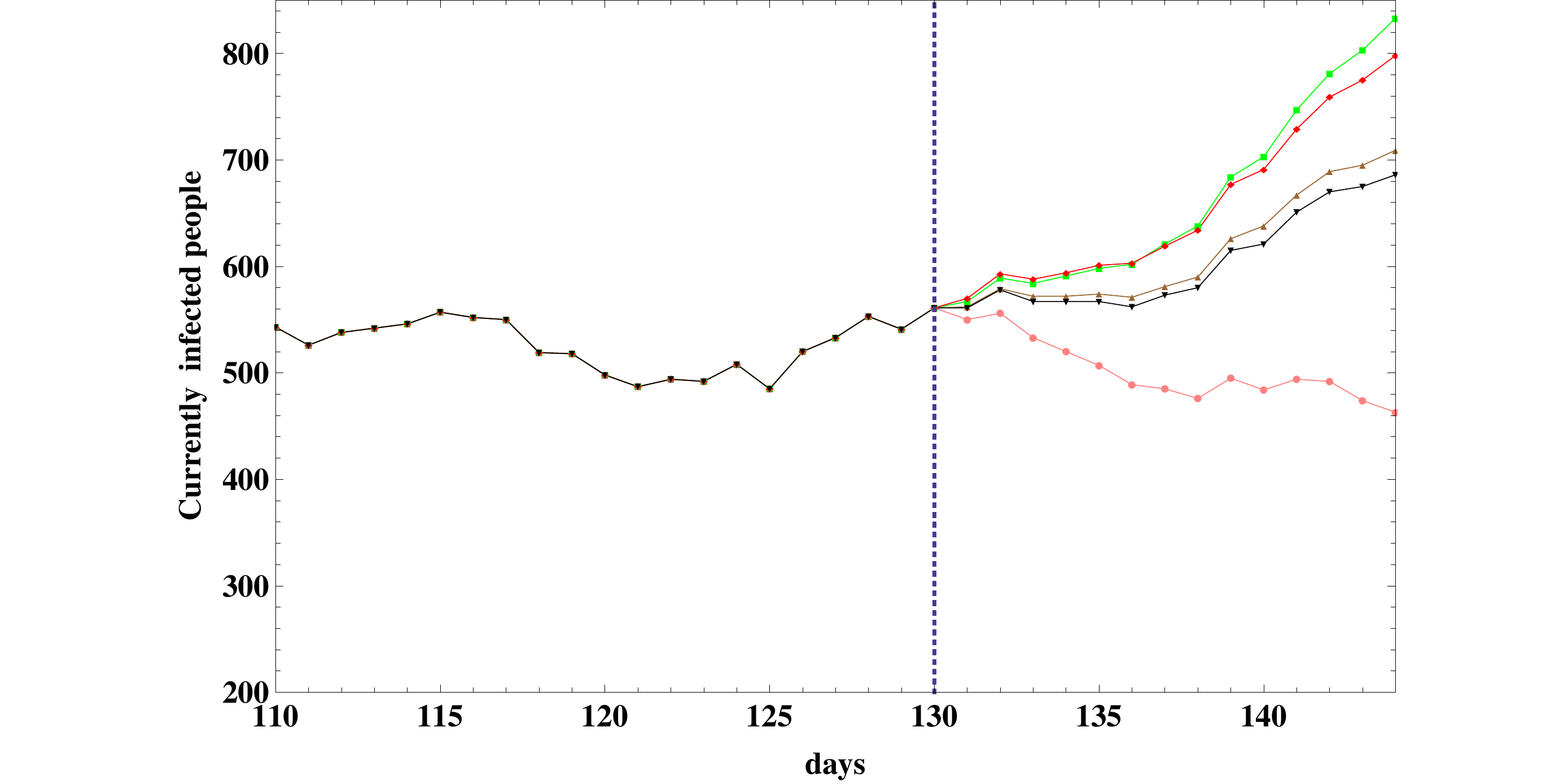}
\caption{Number of currently infected people according to the RI model. The colored lines represent the progression of the number of currently infected people in the next 14 days according to five possible scenarios. The lower pink line corresponds to an absolute decrease of $2 \%$ in the current rate of infection. The middle black and brown lines almost coincide and they correspond to the continuation of the current geometric and arithmetic average rate of infection. The red line represents a progression with a rate corresponding to the maximal attained daily rate of the last 14 days, while the upper green line corresponds to an absolute increase of $1 \%$ in current average rate.} \label{fig:6}
\end{center}
\end{figure}

\begin{enumerate}
\item We determine the arithmetic and the geometric means of the last 14
days and the corresponding future forecasts are plotted in brown and
black respectively. It is clear that with the rate infection registered in Lebanon, and with more people recovered, the number of
currently infected people would slowly increase during the upcoming couple
of weeks, and the geometric and arithmetic means considered lead to very similar forecasts.
\item The infection rate decreases by an absolute value of $2 \%$  and the number of recovered is
also on the rise; hence, the total number of active cases decreases faster.
\item The rate of progression is defined as the maximum of the rates of increase
recorded in the past fourteen days
\begin{align*}
 G_{\text{max}}=\text{Max\ensuremath{\left\{ G_{g_{\ i-(m-1)}},G_{g_{\ i-(m-2)}},....G_{g_{\ i}}\right\} }}
\end{align*}
The number of active cases would witness a steep increase despite of recoveries and deaths.
\item The mean rate of infection increases by an absolute rate of $1 \%$, and the number of currently infected people would rise quickly despite recoveries or deaths from previous cases.
\end{enumerate}

Similarly to the forecasts obtained using the STEIR model, the different scenarios are contingent on the imposed governmental measures as well as on public behavior and social distancing. If precautions are diminished, the number of infections will be on the rise again according to scenarios 3 or 4. The continued enforcement of measures would lead to scenario 1 or more optimistically scenario 2 in case of severe SDM and perhaps the resumption of a countrywide lockdown (Figure  \ref{fig:6}).

\begin{figure}[t]
\hspace*{-1cm}
\begin{center}
\includegraphics[scale=0.5]{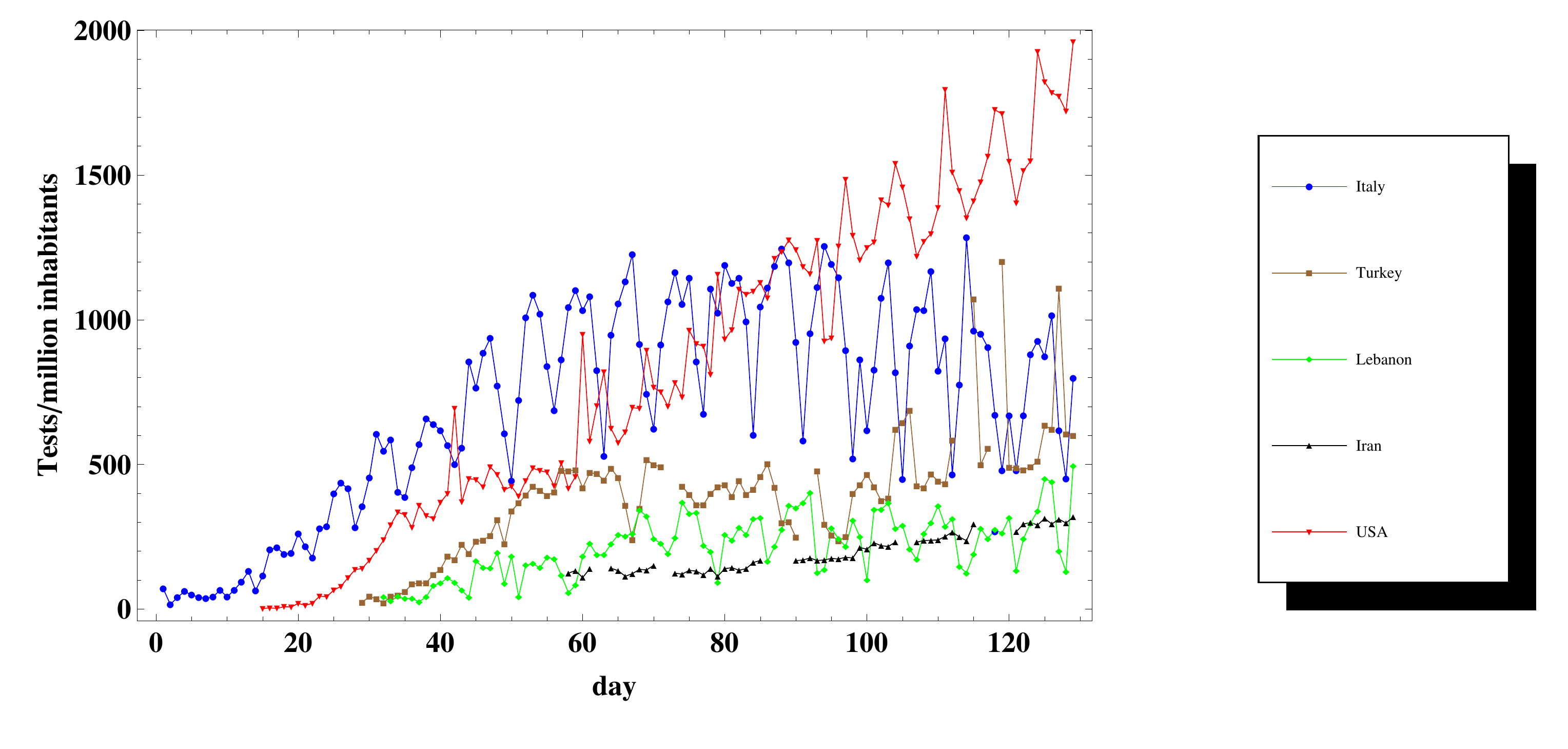}
\caption{The figure shows the number of daily tests conducted per 1 million inhabitants for each of Italy (Blue), Turkey (Brown), Lebanon (Green), Iran (Black) and USA (Red). It compares the daily tests in Lebanon with other countries on the regional and global level.}
\label{fig:7}
\end{center}
\end{figure}

\section{Discussions}

The low infection rate recorded in the second month of the COVID-19 spread in Lebanon could likely be attributed to public
commitment to social distancing and the strict mitigation measures imposed. 
Another factor that could have affected this outcome is the number and extent of distribution of tests undertaken.

To address this issue, we compare the number of daily tests conducted per million residents in Lebanon, with that of  several other countries that witnessed stronger spreads like Italy, USA, Iran and Turkey.
We find that the percentage of people tested in Lebanon is similar
to that in Iran, and tends to rise with time but it is significantly less than the other countries as shown in Figure
\ref{fig:7}. This suggests that the number of detected infections is possibly lower than the actual number of cases. 

Furthermore, we check the ratio of the cumulative number of infections
with respect to the cumulative numbers of tests conducted. This criterion would give the rate of infected out those of tested, hence eliminates the uncertainty related to under-testing. An examination of the publically available data reveals
that the cumulative rate of infection among those who were tested in Lebanon
was $1.37\%$ on June 24th, 2020, compared to $8.17\%$ in the USA, $4.63\%$
in Italy, $6.21\%$ in Turkey and $14.04\%$ in Iran \cite{worldometer2020}.
This is an assertion that the low rate of infection is more related
to the mitigation measures applied in the country, as rates in more affected countries are larger by several orders of magnitude.

On the other hand, the relaxation of measures and the rise of the reproductive rate of infections
due to increased interaction could reinstate a high rate of infections within a couple of weeks as forecasted in Figure \ref{fig:4} and  Figure \ref{fig:6}. This is confirmed by the simulations conducted based both on the STEIR and the RI models. The aforementioned models both forecast a resumption of a quick spread of the disease and an increase in the number of infected people once the SDM are reduced or abandoned, and once the inflow of travelers commences and increases.
We note that the RI model, by construction, already accounts for all infected cases, whether residents or travellers, but it only allows a projection for a short term future behavior of infectious spread.

Our data also depicts that the mean age of infected individuals has fallen from 43.8 years to 36.7~years old between April 8 until June 30, 2020, which might have been generated by the travel inflows during this period.

Under all circumstances, the continued SDM and travel restrictions are essential to keep COVID-19 under control \cite{Perspective2020} until the introduction of effective medications or vaccines, which is estimated to take at least between 12 to 18 months \cite{NG2020}, despite ongoing medical and clinical research around the globe \cite{Vaccines3}.

\section{Conclusion}

This paper presented two different models used in the simulation
of the spread of infectious diseases which are the STEIR model and the
RI model.
We developed the STEIR model, a novel parameterization of the reproductive rate of infection, an improved RI model and adjusted all relevant parameters to fit the available data from
Lebanon. We analyzed in detail the current spread and different forecasts
for potential progressions. We found out that the rate of infection and
the number of infected people fell down quickly due to the rapid fall
in the reproductive number $R$ due to strong mitigation measures.
However, relaxing the measures, partially lifting the travel ban and the resumption of social activity
and interaction would swiftly put the infections on a rapid rise again, thus
reversing the temporary success in limiting the spread of COVID-19.

The results obtained can be further extended in different directions. At the theoretical level, the suggested models can be enhanced and generalized to include other within and cross-country features of the progression of pandemics. Furthermore, the findings can be exploited to draw policies and establish appropriate measures to limit the spread of the disease.

\section*{Acknowledgements}
The authors have no acknowledgements.

\section*{Conflict of interests}
The authors declare that they have no conflict of interests.

\section*{Data availability}
All data can be supplied by the authors upon request.

\end{document}